\documentstyle[aaspp4]{article}

\catcode`\@=11 
\def\@versim#1#2{\vcenter{\offinterlineskip
        \ialign{$\m@th#1\hfil##\hfil$\crcr#2\crcr\sim\crcr } }}
\newcommand{\Ref}{\hangindent=20pt \hangafter=1 \noindent}
\newcommand{\StartRef}{\hyphenpenalty=10000 \raggedright}

\newcommand{\beq}{\begin{equation}}
\newcommand{\eeq}{\end{equation}}
\def\lsim{\mathrel{\mathpalette\@versim<}}
\def\gsim{\mathrel{\mathpalette\@versim>}}

\def\@versim#1#2{\vcenter{\offinterlineskip
        \ialign{$\m@th#1\hfil##\hfil$\crcr#2\crcr\sim\crcr } }}
\catcode`\@=12 

\begin{document}

\title{On the Structure of Advective Accretion Disks At High Luminosity}

\author{Ioulia V. Artemova, Gennadi S. Bisnovatyi-Kogan}
\affil{Space Research Institute, Profsoyuznaya 84/32, Moscow 117810, Russia;
gkogan@mx.iki.rssi.ru, julia@mx.iki.rssi.ru}

\author{Igor V. Igumenshchev\altaffilmark{1}}
\affil{Institute of Theoretical Physics, G\"oteborg University and
Chalmers University of Technology, 412 96 G\"oteborg, Sweden; 
ivi@fy.chalmers.se}

\author{Igor D. Novikov\altaffilmark{2,3,4}}
\affil{Copenhagen University Observatory, Juliane Maries Vej 30,
                DK-2100 Copenhagen, Denmark; novikov@tac.dk}

\altaffiltext{1}{On leave from Institute of Astronomy,
48 Pyatnitskaya St., 109017 Moscow, Russia}
\altaffiltext{2}{Theoretical Astrophysics Center, Juliane Maries Vej 30,
                DK-2100 Copenhagen, Denmark}
\altaffiltext{3}{Astro-Space Center of P.N. Lebedev Physical Institute,
                Profsouznaya 84/32, Moscow, Russia}
\altaffiltext{4}{NORDITA, Blegdamsvej 17, DK-2100 Copenhagen, Denmark}

\medskip
\setcounter{footnote}{0}

\begin{abstract}

Global solutions of optically thick advective accretion disks
around black holes
are constructed. The solutions are obtained by solving numerically a set
of ordinary differential equations corresponding to a steady
axisymmetric 
geometrically thin disk.
We pay special attention to consistently satisfy the regularity conditions
at singular points of the equations. For this reason we analytically
expand a solution at the singular point, and use coefficients of the
expansion in our iterative numerical procedure.
We obtain consistent transonic solutions
in a wide range of values of the viscosity parameter $\alpha$ and
mass accretion rate. We compare two different form of viscosity:
one takes the shear stress to be proportional to the pressure, while
the other uses the angular velocity gradient-dependent stress.

We find that there are two singular points in solutions corresponding to
the pressure-proportional
shear stress. The inner singular point
locates close to the last stable orbit around black hole.
This point changes its type from a saddle to node depending on values of
$\alpha$ and accretion rate.
The outer singular point
locates at larger radius and is always of a saddle-type. We argue that,
contrary to the previous investigations,
a nodal-type inner singular point does not introduce multiple solutions.
Only one integral curve, which corresponds to the unique global solution,
passes simultaneously the inner and outer singular points independently of
the type of inner singular point.
Solutions with the angular velocity gradient-dependent shear stress
have one singular point which is always of a saddle-type and corresponds
to the unique global solution.
The structure of accretion disks corresponding to both viscosities
are similar.

\noindent {\em Subject headings:} Accretion, accretion disks ---
black hole physics --- hydrodynamics
\end{abstract}


\section{Introduction}

Accretion discs are formed when the matter with a large angular momentum
is falling into a black hole or another gravitating body.
The well known objects  where the accretion disks are found are
protostar nebulae, binary
X-ray sources, 
cataclysmic 
variables, active galactic nuclei and others.
In this paper we discuss accretion disks around black holes.
The standard model of geometrically thin
accretion disk has been developed by Shakura (1972),
Novikov \& Thorne (1973) and
Shakura \& Sunyaev (1973), and has played a significant role
in the development of accretion theory (see Pringle 1981;
Frank, King, \& Rain 1992; Kato, Mineshige, \& Fukue 1998 for reviews).
The standard model bases on the 
vertically averaged approach to equilibrium,
and a
suggestion of the local thermal balance in which the viscous heating
of the gas is balanced by the local radiative cooling. Non-local effects,
like the radial advection of thermal energy
and the transonic nature of accretion flow,
are neglected in the standard model.
This simplified approach allows to reduce the general problem
to a set of algebraical equations.
Such a simple description becomes possible due to an approximate
parameterization of the viscosity stress tensor with one non-zero component,
$$
t_{r\phi}=-\alpha P, \eqno(1)
$$
suggested by Shakura (1972).
The standard model gives a satisfactory appropriate solution of the problem
at low accretion rates $\dot{M}\la 16 L_{Edd}/c^2$, where
$L_{Edd}$ is the Eddington luminosity.

Simplified solution with inclusion of the advective terms into equations described
the vertically integrated models of accretion disks was obtained by
Paczy\'nski \& Bisnovatyi-Kogan (1981).
This approach with some modifications have been used
by many researchers to study transonic accretion flows
around black holes (Muchotrzeb \& Paczy\'nski 1982;
Muchotrzeb 1983;
Matsumoto et al. 1984; Fukue 1987; Abramowicz et al. 1988;
Chen \& Taam 1993; Beloborodov 1998).
The importance of the transonic nature of the accretion flows
on the disk structure has been emphasized
by H\"oshi \& Shibazaki (1977), Liang \& Thompson (1980) and
Abramowicz \& Zurek (1981), and later studied in more details
by Abramowicz \& Kato (1989).


Despite a significant progress in the study of
optically thick accretion disks obtained during almost three decades
there are a number of unsolved problems still posed in the theory.
The problems are connected with a possible non-uniqueness
of a solution at $\alpha\ga 0.01$
and a non-standard behavior of a singular point type.
It was reported by Matsumoto et al. (1984),
Muchotrzeb-Czerny (1986) and Abramowicz et al. (1988)
that in the case of viscosity prescription (1)
the singular point changes its type
from a saddle to node when one increases $\alpha$.
The presence of the nodal-type singular point leads to creating of
a possibility of multiple solutions as the authors have claimed.
A similar change of the singular point type was reported by Chen \& Taam
(1993), who used the angular velocity gradient-dependent viscous stress,
$$
t_{r\phi}=\rho\nu r {d\Omega\over dr}, \eqno(2)
$$
where $\nu$ is the kinematic viscosity coefficient defined by (14).
Narayan, Kato, \& Honma (1997) have compared two forms of viscosity
(1) and (2) in the case of radiatively inefficient advection-dominated
accretion flows. They concluded that the structures of flows corresponded
to both viscosities are similar at $\alpha<0.15$.

In this paper we show that the mentioned
problems have been created by several inconsistencies
in the preceding studies. Some problems are
connected with an inaccurate averaging
of the equations over a disk thickness
(Chen \& Taam 1993),
another ones appear due to an incomplete investigation of
the singular points (Abramowicz et al. 1988).
We have found  that in the case of viscosity prescription (1)
a set of equations describing the vertically averaged advective
accretion disks has
{\it two} singular
points, independed of $\alpha$ and accretion rate.
Note, that multiplicity of singular points in solutions for accretion flows
in Paczy\'nski-Wiita potential (3) was revealed by Fukue (1987),
Chakrabarti \& Molteni (1993) and Chakrabarti (1996) in a somewhat different
context. We have shown, that at $\alpha\la 0.01$ the inner and outer
(with respect to the black hole location)
singular points are of the saddle type, and only one
integral curve (``separatrix'') which crosses
the inner point simultaneously crosses
the outer one.
This separatrix corresponds to the unique global solution which is
determined by two parameters, $\alpha$ and $\dot{m}=\dot{M}c^2/L_{Edd}$,
for a given black hole mass.
In Figure~1a the structure of integral curves is schematically
represented in the vicinity of the global solution which is shown by the
thick line.
At larger $\alpha\ga 0.1$
the inner singular point is changed to a nodal-type one,
while the outer point remains of a saddle-type.
There is still one integral curve which goes continuously through
both singular points providing a unique global solution,
as it is shown in Figure~1b.

In the case of viscosity prescription (2) we have found that
there is only one singular point which is always a saddle,
and only one physical solution which passes through this point exists.
Solutions which correspond to both forms of viscosity (1) and (2)
are very close at low $\alpha$ limit, $\alpha\la 0.1$.

We have developed a numerical method to solve the set of
equations describing the vertically averaged
advective accretion disks.
The method is based on the standard relaxation technique and
explicitly uses conditions at the inner singular point and its vicinity.
We have obtained these conditions by
expanding a solution into power series around
the singular point.
Such a modification of the method allows
to construct solutions which
smoothly pass the singular points and satisfy the regularity
conditions at these points
with high computer precision in wide range of parameters
$\alpha$ and $\dot{m}$.

The paper is organized as follows. In \S2 we formulate a
mathematical approach to the problem, write a set of equations, and
formulate boundary conditions. In \S3 we investigate  critical
points and discuss behavior of physical values in their vicinity.
In \S4 we describe our numerical results and discuss
them in \S5.
Details of the numerical method and explicit expansion of physical
quantities in the vicinity of the critical points are represented
in Appendixes~A and B, respectively.

\section{Problem formulation}

We consider a steady geometrically thin accretion disk around a non-rotating
black hole.
For simplicity, we use the pseudo-Newtonian approach to describe
the disk structure in 
the vicinity of a black hole.
In the approach the general relativistic
effects are simulated by using Paczy\'nski-Wiita potential
(Paczy\'nski \& Wiita 1980)
$$
\Phi(r)=-{GM\over r-2r_g}, \eqno(3)
$$
where $M$ is the black hole mass and 
$2r_g=2GM/c^2$
is the gravitational radius.
The disk self-gravity is neglected.

A general problem of investigation of two-dimensional structure of the
accretion disks (in the radial and vertical directions) can be reduced
to a one-dimension problem
by averaging the disk structure in the vertical direction.
In this formulation equations which are described the radial disk structure
are written for the midplane density $\rho$, pressure $P$, radial velocity
$v$ and angular velocity $\Omega$.
The mass conservation equation takes the form,
$$
\dot{M}=4\pi r h \rho v, \eqno(4)
$$
where $\dot{M}$ is the accretion rate, $\dot{M}>0$,
and $h$ is the disk half-thickness,
which is expressed in terms of the isothermal sound
speed $c_s=\sqrt{P/\rho}$ of  gas,
$$
h={c_s\over\Omega_K}. \eqno(5)
$$
The equations of motion in the radial and azimuthal directions are
$$
v{dv\over dr}=-{1\over\rho}{dP\over dr}+(\Omega^2-\Omega^2_K)r, \eqno(6)
$$
$$
{\dot{M}\over 4\pi}{d\ell\over dr}+{d\over dr}(r^2 h t_{r\phi})=0,
\eqno(7)
$$
where $\Omega_K$ is the Keplerian angular velocity,
$\Omega_K^2=GM/r(r-2r_g)^2$,
$\ell=\Omega r^2$ is the specific angular momentum and $t_{r\phi}$ is the
($r$, $\phi$)-component of the viscous stress tensor.
Other components of the stress tensor are assumed to be negligiblly small.

The vertically averaged energy conservation equation can be written
in the form\footnote{The vertical averaging in equation (8) have been done in
different way by different authors (compare e.g.
Shakura \& Sunyaev 1973, and Abramowicz et al. 1988). Our choice of
coefficients in (9)-(11), 
following Chen \& Taam (1993) may be not the optimal one.
Aposteriory analysis had shown that using 4 in the denominator of (9)
instead of 2, would be a more consistent choice, but this change has a
little influence on our numerical results.},
$$
F_{adv}=F^{+}-F^{-}, \eqno(8)
$$
where
$$
F_{adv}=-{\dot{M}\over 2\pi r}\left[{dE\over dr}+P{d\over dr}
\left({1\over\rho}\right)\right], \eqno(9)
$$
$$
F^{+}= h t_{r\phi}r{d\Omega\over dr}, \eqno(10)
$$
$$
F^{-}={2 a T^4 c \over 3 \kappa \rho h}, \eqno(11)
$$
are the advective energy flux, the viscous dissipation rate and
the cooling rate per unit surface, respectively,
$T$ is the midplane temperature,
$\kappa$ is the opacity and $a$ is the radiation constant.

The equation of state for accretion matter consisted of
a gas-radiation mixture is
$$
c_s^2={\cal R}T+{1\over 3}{aT^4\over\rho},  \eqno(12)
$$
where ${\cal R}$ is the gas constant.
The specific energy of the mixture is
$$
E={3\over 2}{\cal R}T+{aT^4\over\rho}. \eqno(13)
$$

We consider two prescriptions of viscosity in our models.
In one case we adopt a simple relation (1) between the viscous stress
and pressure.
In another case we assume the angular velocity gradient-dependent viscous
stress (2),
where the viscosity $\nu$ is taken in the form
$$
\nu={2\over 3}\alpha c_s h. \eqno(14)
$$
In the limit $\Omega\longrightarrow\Omega_K$ both prescriptions (1) and (2)
coincide.

Integrating equation (7) we obtain
$$
r^2 h t_{r\phi}=-{\dot{M}\over 4\pi}(\ell-\ell_{in}), \eqno(15)
$$
where $\ell_{in}$ is an integration constant and has a meaning
of the specific angular momentum of accreting matter at the black hole horizon.
Depending on the used viscosity prescription (1) or (2) expression (15)
results in
an algebraical equation or a first order differential equation, respectively.
So, in the case of viscosity prescription (1) we have only two
first order differential equations (6) and (8),
which require to set two parameters as boundary conditions.
In the case of viscosity prescription (2) we obtain additional differential
equation from (15) and have to set three parameters as boundary conditions.
The integration constant $\ell_{in}$ is chosen to obtain a global transonic
solution with a subsonic part at large radii and a supersonic part
close to the black hole horizon.

\section{Investigation of singular points}

\subsection{$\alpha P$ viscosity prescription}

We consider first the case of viscosity prescription (1).
From (15) we obtain the algebraical expression for $\Omega$,
$$
\Omega={\ell_{in}\over r^2}+\alpha{c_s^2\over v r}. \eqno(16)
$$
Using (16) the
system of equations (6) and (8) can be reduced to the following form,
$$
r{v'\over v}={N_1\over D_1}, \eqno(17)
$$
$$
r{c_s'\over c_s}=(1-{\cal M}^2){N_1\over D_1}+
1-r{\Omega_K'\over\Omega_K}+
{\Omega^2-\Omega_K^2\over c_s^2}r^2, \eqno(18)
$$
where
$$
N_1=\left(1-r{\Omega_K'\over\Omega_K}+
{\Omega^2-\Omega_K^2\over c_s^2}r^2\right)
\left(7-{3\over 2}\beta{1+\beta\over 4-3\beta}-{\alpha^2\over{\cal M}^2}
\right)+
\left(1-r{\Omega_K'\over\Omega_K}\right)
\left(1+{3\over 2}\beta{1-\beta\over 4-3\beta}\right)+
$$
$$
\alpha{\ell_{in}\over vr}+{1\over 2}{\alpha^2\over{\cal M}^2}-
{1-\beta\over\dot{m}}{\Omega_K r^2\over c_s r_g}, \eqno(19)
$$
$$
D_1=({\cal M}^2 - 1)
\left(7-{3\over 2}\beta{1+\beta\over 4-3\beta}-{\alpha^2\over{\cal M}^2}
\right) -
\left(1+{3\over 2}\beta{1-\beta\over 4-3\beta}+
{1\over 2}{\alpha^2\over{\cal M}^2}\right). \eqno(20)
$$
In equations (17)-(20) we use the following notations:
$v'\equiv dv/dr$, $c_s'\equiv dc_s/dr$, $\Omega_K'\equiv d\Omega_K/dr$,
$\beta={\cal R}T/c_s^2$ and
${\cal M}=v/c_s$.
From (4) and (16) it follows the algebraical equation for $\beta$,
$$
\beta^4-(1-\beta){3\over 4\pi}{\dot{M}\Omega_K{\cal R}^4\over a r v c_s^7}=0.
\eqno(21)
$$

The equation $D_1=0$ defines singular points of
the differential equations (17) and (18), and can be reduced to the
following form,
$$
\left(7-{3\over 2}\beta{1+\beta\over 4-3\beta}\right){\cal M}^4-
\left(\alpha^2+8-{3\beta^2\over 4-3\beta}\right){\cal M}^2+
{\alpha^2\over 2}=0. \eqno(22)
$$
Equation (22) is a quadratic equation with respect to ${\cal M}^2$
and has two positive roots, which
correspond to two singular points:
$$
{\cal M}_{1,2}^2={1\over 2}\left[\alpha^2+8-{3\beta_s^2\over 4-3\beta_s}\pm
\sqrt{\left(\alpha^2+8-{3\beta_s^2\over 4-3\beta_s}\right)^2-
2\alpha^2\left(7-{3\over 2}\beta_s{1+\beta_s\over 4-3\beta_s}\right)}\right]
\left(7-{3\over 2}\beta_s{1+\beta_s\over 4-3\beta_s}
\right)^{-1}, \eqno(23)
$$
where $\beta_s$ is the value of $\beta$ taken at the singular point.

In $\alpha\ll 1$ limit we have:
$$
{\cal M}_{1}^2=\left(8-{3\beta_s^2\over 4-3\beta_s}\right)
\left(7-{3\over 2}\beta_s{1+\beta_s\over 4-3\beta_s}\right)^{-1} \eqno(24)
$$
and
$$
{\cal M}_{2}^2={\alpha^2\over 2}
\left(8-{3\beta_s^2\over 4-3\beta_s}\right)^{-2}. \eqno(25)
$$
The first singular point, in which ${\cal M}_s={\cal M}_1$, locates
close to the
black hole last stable orbit at $r=6 r_g$. The corresponding values of
${\cal M}_1$ are $1.118$ and $1.069$ for the gas pressure supported
($\beta=1$) and the
radiation pressure supported ($\beta=0$) accretion flows, respectively.
This point is an analogy of the singular point in a spherical flow, where
the point divides the subsonic and supersonic regions of accretion flow.
The second singular point, in which ${\cal M}_s={\cal M}_2$,
located at larger radius, is the result of simplified
viscosity prescription (1).
We will use the notations $(r_s)_{in}$ and $(r_s)_{out}$ for positions
of the inner and outer singular points, respectively.

At the singular points the numerator $N_1$ and denominator $D_1$
must simultaneously vanish to provide
a regular behavior for a global solution.
The type of the singular points must be consistent with a
transonic nature of solution. For example, a spiral-type singular point
does not satisfy the latter requirement, but a saddle or nodal-type
point does it. Detailed analysis of topology in vicinity of singular points
was done for thin accretion disks
under isothermal approximation by Abramowicz \& Kato (1989).
They showed that saddle, nodal or spiral types
are formally possible, but only saddle and nodal points are physically
relevant. This study had confirmed  
the previously obtained numerical results
by Matsumoto et al. (1984). The latter authors demonstrated in a framework
of the isothermal accretion disks that the type of the inner singular
point is defined by value of $\ell_{in}$. At larger $\ell_{in}$ the point
is a spiral, at smaller $\ell_{in}$ there is no inner singular
point at all, and only unique choice of $\ell_{in}$ of moderate values
corresponds to a saddle or nodal-type singular point.
The choice between saddle or nodal-type
singular points can be done only by constructing a global model of the disk.

\subsection{$\Omega$-gradient-dependent viscous stress}

In the case of viscosity prescription (2) the
differential equations (6), (8) and
(15) can be reduced to the following form,
$$
r{\Omega'\over\Omega}=-{3\over 2}{\Omega_K r v\over\alpha c_s^2}
\left(1-{\ell_{in}\over\Omega r^2}\right), \eqno(26)
$$
$$
r{v'\over v}={N_2\over D_2}, \eqno(27)
$$
$$
r{c_s'\over c_s}=(1-{\cal M}^2){N_2\over D_2}+
1-r{\Omega_K'\over\Omega_K}+
{\Omega^2-\Omega_K^2\over c_s^2}r^2, \eqno(28)
$$
where $\Omega'\equiv d\Omega/dx$, $\beta$ is defined by (21) and
$$
N_2=
\left(1-r{\Omega_K'\over\Omega_K}+
{\Omega^2-\Omega_K^2\over c_s^2}r^2\right)
\left(7-{3\over 2}\beta{1+\beta\over 4-3\beta}\right)+
{3\over 4}{\Omega^2\Omega_K r^3 v\over\alpha c_s^4}
\left(1-{\ell_{in}\over\Omega r^2}\right)^2+
$$
$$
\left(1-r{\Omega_K'\over\Omega_K}\right)
\left(1+{3\over 2}\beta{1-\beta\over 4-3\beta}\right)-
{1-\beta\over\dot{m}}{\Omega_K r^2\over c_s r_g}, \eqno(29)
$$
$$
D_2=({\cal M}^2-1)\left(7-{3\over 2}\beta{1+\beta\over 4-3\beta}\right)-
\left(1+{3\over 2}\beta{1-\beta\over 4-3\beta}\right). \eqno(30)
$$

There is only one singular point of equations (26)-(28) defined by
the equation $D_2=0$.
The point is an analogy to the inner singular point discussed in \S3.1.
To be consistent with \S3.1 we use notation $(r_s)_{in}$ for the position
of the point. At $(r_s)_{in}$  we have
$$
{\cal M}_s^2=\left(8-{3\beta_s^2\over 4-3\beta_s}\right)
\left(7-{3\over 2}\beta_s{1+\beta_s\over 4-3\beta_s}\right)^{-1}. \eqno(31)
$$
Note, that the expression for ${\cal M}_1$ given by (24) coincides with
(31). The latter could mean that the properties of global solutions
in the case of viscosity prescriptions (1) and (2)
are very similar in the inner
part of flow at the limit of small viscosity, $\alpha\ll 1$.
Our numerical results confirm this conclusion.

Abramowicz \& Kato (1989) studied analytically the type of singular point
in the isothermal disks in the case of viscosity prescription (2).
They showed that the point is always a saddle, and
there is no case of a node. This conclusion differs from one obtained
in the case of viscosity prescription (1). Our numerical models
confirm this dependence of the singular point type on a form of viscosity.

\section{Numerical results}

To be used in the numerical method
the sets of differential equations (17)-(18)
and (26)-(28) have been re-written in the dimensionless form
using the following dimensionless quantities:
$\tilde{r}=r/r_g$, $\tilde{v}=v/c$, $\tilde{c}_s=c_s/c$,
$\tilde{\Omega}=\Omega r_g/c$, $\tilde{\ell}=\ell/r_g c$, $m=M/M_\odot$.
In the subsequent discussions we will use these dimensionless quantities
skipping in the notations the `tilde' mark.
The used method is described in Appendix~A.
We have calculated a number of models varied by
the viscosity prescriptions and parameters $\alpha$ and
$\dot{m}$.
The black hole mass $m$ contributes into the dimensionless equations
in the combinations
$$
D={ac^4\kappa GM_\odot\over {\cal R}^4}{m\over\dot{m}}.
$$
The parameter $D$ was taken to be inversely proportional to $\dot{m}$
with $m=10$, ${\cal R}=1.65\cdot 10^8\,erg\,g^{-1}K^{-1}$ and
$\kappa=0.4 \,cm^2g^{-1}$.
The numerical grid covers the radial range form $r_{in}$
located at the inner singular point, $r_{in}=(r_s)_{in}$,
till $r_{out}\approx 10^4 r_g$.

We discuss first the influence of the numerical outer
boundary conditions on our models. We have found that the models
are {\it insensitive} to the specific values of
the outer boundary conditions.
By fixing $\alpha$ and $\dot{m}$ the unique
global transonic solution is fully determined. This solution also uniquely
determines the outer boundary values: two or three values
depending on the used viscosity prescription (1) or (2), respectively.
In general, we do not know a priori these `correct' boundary values
which the global solution passes through, and
consequently, our assumed numerical boundary values are quite arbitrary.
But, this values should be close enough to the `correct' ones
due to reason of numerical stability.
Also, the `correct' outer values are very close, but not exactly
equal, to the
values obtained from the standard model for the Keplerian accretion disks.
Calculations show that all numerical solutions which have the same
$\alpha$ and $\dot{m}$, but different boundary values at different $r_{out}$,
converge to the `common' solution which is not affected by
the outer boundary. This `common' solution represents
the global solution which we seek.
Significant differences between some numerical
solution and the `common' one (with relative errors $\ga 10^{-4}$)
are observed only in $2-3$ grid points
before the last outer point at $r_{out}$.
Such a behaviour of the numerical
solutions can be explained by
special properties of difference equations (A1)
at $\varepsilon\approx 1$.
At small $\varepsilon$ the method is unstable.

Each model is characterized by value of $\ell_{in}$ [see eq.(15)]
which has a sense of a   
specific angular moment of matter
infalling into black hole.
Figure~2 [panels~(a) and (b)] shows the dependence of $\ell_{in}$ on
accretion rate $\dot{m}$ for three values of $\alpha=0.01$, $0.1$ and $0.5$,
and two forms of viscosity prescription (1) and (2).
At low $\dot{m}\la 1$ the value of $\ell_{in}$ is independent of $\dot{m}$,
but weakly varies with $\alpha$.
In the low $\alpha$ case, $\alpha=0.01$ and $0.1$, the values of $\ell_{in}$
are close to the minimum value of
the Keplerian angular momentum, $(\ell_K)_{min}=3.6742$.
At high $\dot{m}\ga 0.1$ the values of
$\ell_{in}$ deviate from $(\ell_K)_{min}$
to larger or smaller values depending on $\alpha$.
In the case $\alpha=0.01$ and $0.1$
one can see only minor differences between models with different forms
of viscosity.
But, for large $\alpha=0.5$ the difference in values of $\ell_{in}$
increases. Unfortunately, we 
had been able to calculated only a limited number
of models in the case of viscosity prescription (2) due to technical reason
(see Appendix~A for details), and our comparison of both prescriptions
is not complete in this respect.

Figure~2 [panels~(c) and (d)]
shows locations of the inner singular points $(r_s)_{in}$
as a function of $\dot{m}$
for different values of $\alpha$ and two different viscosity prescriptions.
Similar to the case of $\ell_{in}$ the models at low $\dot{m}$
show a weak dependence of $(r_s)_{in}$ on $\dot{m}$.
In the low $\alpha$ models (squares and circles in Figure~2)
the values of $(r_s)_{in}$ are close
to the location of the black hole last stable orbit at $r=6$.
At high $\dot{m}\ga 0.1$ the values of $(r_s)_{in}$
are decreasing functions of $\dot{m}$ in the case
of low $\alpha=0.01$ and $0.1$, and non-monotonically behave
in the case of $\alpha=0.5$ (triangles in Figure~2).

Figure~3 shows locations of the outer singular points $(r_s)_{out}$
in the case of viscosity prescription (1)
as a function of $\dot{m}$ for different values of $\alpha$.
Values of $(r_s)_{out}$ are an increasing function of $\dot{m}$ and
show a power-law behaviour at $\dot{m}\ga 3$.
It is interesting to note that values of $(r_s)_{out}$ are almost
independent of $\alpha$.

Examples of the specific angular momentum distribution $\ell(r)$
are shown in Figure~4 for $\dot{m}=160$ and three values of
$\alpha=0.01$ (short-dashed line), $0.1$ (solid line) and
$0.5$ (dotted line). The distributions correspond
to viscosity prescription (1).
The location of the inner singular points are indicated by
the correspondent points on the curves.
The Keplerian angular momentum is displayed by the
long-dashed line for comparison.
Only the low viscosity model with $\alpha=0.01$ has a super-Keplerian part
in $\ell(r)$. Models with larger viscosity are everywhere sub-Keplerian.
Note, that the singular point in the low viscosity model (short-dashed line)
locates in the inner sub-Keplerian part of the disk.

Figure~5 shows the dependence of $\beta_s$ on $\dot{m}$
at the inner singular points.
The change of value of
$\beta$ form 1 to 0 corresponds to the change of a state from the gas pressure
to radiative pressure dominated one.
The thin disks with $\beta\simeq 1$ are locally stable,
whereas the parts of the disk in which $\beta\simeq 0$
are thermally and viscously unstable at $\dot{m}\la 100$
(Pringle, Rees, \& Pacholczyk 1973). At larger $\dot{m}\ga 100$
the instability can be suppressed by the advection effect
(Abramowicz et al. 1988).
We have found a weak dependence of $\beta_s(\dot{m})$ on the assumed viscosity
prescriptions.

Using analysis discussed in Appendix~B we have determined a type of
singular points in our numerical solutions.
In Figures~2-5 the saddle-type points are indicated by the solid squares,
circles and triangles.
The nodal-type points are represented by
the 
corresponding 
empty dots in the same figures.
In the case of viscosity prescription (1)
the solutions have two singular points (see \S3.1).
The inner singular points, $(r_s)_{in}$, can be saddles
or nodes depending on values of $\alpha$ and $\dot{m}$.
Note that the change of type from a saddle to nodal one
does not introduce any features in solutions.
The outer singular points, $(r_s)_{out}$, are always of a saddle-type.
In the case of viscosity prescription (2)
the solutions have only inner singular points
(see \S3.2) which are always of a saddle-type.

In models with low $\dot{m}\la 16$ and low $\alpha\la 0.1$
the values of $r_s$ and $\ell_{in}$ are quantitatively
very close to the last stable orbit  
location ($r_{in}=6$) and
value of $\ell_{in}$ [$\ell_{in}=(\ell_K)_{min}$] assumed in the
standard model (Shakura \& Sunyaev 1973).
The radial structure in our models is also very close
to the one for the standard model
in the same range of $\dot{m}$ and $\alpha$.
Such a good coincidence means that
the advective terms in equations (6) and (8)
are negligiblly small in considered models.
However, the high $\alpha$ models show quite
significant deviation from the standard model independently of
$\dot{m}$ (see Figure~2).

At high accretion rates, $\dot{m}\ga 16$, the effect of advection
becomes significant. 
We illustrate it by calculating the luminosity $L$ of disk
in the case of viscosity prescription (1),
$$
L=4\pi\int_{(r_s)_{in}}^\infty F^- r dr=
2L_{Edd}\int_{(\tilde{r}_s)_{in}}^\infty {(1-\beta)\tilde{c}_s\tilde{r}^{1/2}
\over \tilde{r}-2}d\tilde{r}, \eqno(32)
$$
where $F^-$ is given by (11).
Figure~6 shows calculated dependences of $L/L_{Edd}$ on $\dot{m}$
for different values of $\alpha=0.01$, $0.1$, $0.5$ (short-dashed, solid
and dotted lines, respectively).
There is a simple linear relation $L/L_{Edd}=\eta\dot{m}$ 
in the standard model, in which the advection is neglected.
The radiative efficiency $\eta$ is a constant and equals $\eta=1/16$ 
in the case of gravitational potential (3).
We plot this relation by the straight long-dashed line in Figure~6.
One can clearly see from the figure that effect of advection results in
reduction of luminosities with respect to the one for the 
standard model at $\dot{m}\ga 16$.
There is a weak dependence of the luminosity on value of viscosity in disks.

Finally note, that our numerical solutions
corresponding to viscosity prescription (1)
have some resemblance to
the results of Abramowicz et al. (1988),
but they show important quantitative differences,
especially for large $\alpha$ and $\dot{m}$.

\section{Discussion}

We have obtained unique solutions for structure of advective
accretion disk in a wide range of accretion rates and $\alpha$-parameters.
Both viscosity prescriptions (1) and (2)
have been investigated.
The solutions corresponding to both prescriptions are very close
for $\alpha \la 0.1$, and begin to differ at larger $\alpha$.  This is
connected, probably with larger deviation of the angular velocity $\Omega$
from the Keplerian one, $\Omega_{K}$, leading to larger difference between
$t_{r\phi}$ in both prescriptions. Unfortunately, our 
comparison
of the prescriptions is not complete due to technical problems in calculation
of the high viscosity models in the case of viscosity prescription (2).

The main difference of the present study from the previous ones is
in using more sophisticated numerical technique which accurately treats
the regularity conditions in the inner singular point of equations.
We have performed an analytical expansion at the singular point
to calculate the derivatives of physical quantities. These derivatives
help us to find the proper integral curve 
passing through the singular point.
The approach allow us to avoid numerical instabilities and
inaccuracies, appearing when only variables at the singular point,
but not its derivatives, are included into a numerical scheme.


We have found different behaviour of integral curves depending on
used viscosity prescription. In the case of viscosity prescription (1)
there are two singular points located at $(r_s)_{in}$ and $(r_s)_{out}$.
The inner point, $(r_s)_{in}$, locates close to the last stable
black hole orbit (see Figure~2), and is an analogy of the singular point
in spherical flow, where the point divides the subsonic
and supersonic regions. The location of the outer point, $(r_s)_{out}$,
is determined by the accretion rate (see Figure~3).
At low $\alpha\la 0.1$ both points are of a saddle-type.
Only one integral curve (``separatrix'') simultaneously crosses
two saddle-type points, as it is shown in Figure~1a, and corresponds
to the global solution which smoothly connects the supersonic
innermost region of the accretion disk and the subsonic outer
(formally at $r=\infty$) parts.
For larger $\alpha\ga 0.1$ the inner singular point changes its type
to a node.
There was suggestion by Muchotrzeb-Czerny (1986) and Abramowicz et al. (1988)
that there is no unique solution in this case,
because of all integral curves cross the node.  Existence of a unique
separatrix crossed simultaneously both singular
points preserves a uniqueness of the solution in this case (see Figure~1b).
The conclusion of Muchotrzeb-Czerny (1986) and
Abramowicz et al. (1988) is probably connected
with their neglection of outer singular points inherited to the problem.


Matsumoto et al. (1984) used slightly different form of equations
(6) and (8), and they found that in this case only one singular point exists
and changes type from a saddle to node.
The difference with respect to our results
arises because of using different form of the pressure
gradient force [the first term on the right hand side of equation (6)].
We use the pressure and density taken at the equatorial plane in this term,
whereas Matsumoto et al. (1984) used the vertically averaged quantities in it.
In the latter case the term has the following form,
$$
{1\over\Sigma}{{\cal P}\over dr}, \quad
{\rm where} \quad
\Sigma=2\int_0^h\rho dz \quad {\rm and} \quad
{\cal P}=2\int_0^h P dz. \eqno(33)
$$
The vertically integrated approach (33) introduces the difference
because in this case the free terms with $\alpha^2$ in (20) are absent.
Formally, it corresponds to location of the outer singular point at
infinity, where ${\cal M}_2=0$.
Such a visible difference is not qualitatively
important for the physical solution,
because conditions at the outer singular point are only shifted to
infinity, and the integral curve itself has little changes.
Thus, similar to our results, in the approach by Matsumoto et al. (1984)
the inner critical points of both types, a saddle or node,
correspond to a unique solution.

In the case of viscosity prescription (2) we have found that
one singular point exist.
The point is always of a saddle-type and determines a unique solution.
Another results had been obtaining by Chen \& Taam (1993).
They also found that equations have one singular point, but
the point changes its type from a saddle to nodal one, depending on
$\alpha$ and the accretion rate.
It is not clear why such a result was obtained.
There are two main differences in our equations (6) and (8), and those
used by Chen \& Taam (1993). First, they used the same vertical
averaging for the equation of motion as Matsumoto et al. (1984).
Second, they used the vertically averaged energy equation which
corresponds to the inappropriate polytropic relation,
${\cal P}\propto\Sigma^\gamma$, when one neglects terms
corresponding to the viscous heating and radiative cooling.
Here $\gamma$ is the effective adiabatic index, and other notations are similar
to those used in (33).
Our equation (8) corresponds to the correct polytropic relation,
$P\propto\rho^\gamma$.
It could be that one of the mentioned differences results in the change
of critical point type.

\noindent{\it Acknowledgments.}
This work was supported in part by RFBR through grant 99-02-18180,
the Royal Swedish Academy of Sciences, 
the Danish Natural Science Research Council through grant No 9701841, 
Danmarks Grundforskningsfond through its support for establishment of
the Theoretical Astrophysics Center.

\appendix
\section{Numerical method}

We use the finite-difference method to solve the systems of ordinary
differential equations discussed in $\S$3.
The method has some resemblance to that used by
Igumenshchev, Abramowicz \& Novikov (1998).
In this approach the problem
is reduced to a solution of a system of non-linear
algebraical equations written for a each pair of neighboring numerical
grid points. The numerical grid $\{r_i\}$ extends over about three orders
of magnitude
in the radial direction. We look for a numerical solution in which
the location of the inner grid point $r_1$ coincides with the
location of the singular point $(r_s)_{in}$ near black hole.
In our method we approximate the differential equation
$dy/dr=f(r,y)$ by the following finite differences,
$$
{y_{i}-y_{i-1}\over r_{i}-r_{i-1}}=\varepsilon f_{i-1}-
(1-\varepsilon)f_{i}, \qquad i=1,2,...,I, \eqno(A1)
$$
where the function $y(r)$ must be replaced by $v(r)$, $c_s^2(r)$, and
additionally by $\Omega(r)$
in the case of viscous prescription (2),
$\varepsilon$ is a parameter, $I$ is the number of grid points,
and the lower indices indicate the corresponding grid point.
The value of parameter $\varepsilon$ is chosen to provide a stability of
the numerical scheme. We use $\varepsilon=1$ in most of the cases.

We use the Newton-Raphson iteration scheme to solve the set of equations (A1).
In general formulation equations (A1) can be represented by
$K$ functional relations,
involving $K$ variables $y_k$,
$$
F_k(y_1,y_2,...,y_K)=0,\qquad k=1,2,...,K, \eqno(A2)
$$
or in the vector notation, ${\bf F}({\bf y})=0$. The $(n+1)$-iteration
improvement of an approximate solution ${\bf y}^{n}$ of (A2) has the form,
$$
{\bf y}^{n+1}={\bf y}^{n}+\omega^n\cdot\delta{\bf y}^n, \eqno(A3)
$$
where $\omega^n$ is a parameter, $\omega^n\leq 1$, and
the correction $\delta{\bf y}^n$ is the solution of the matrix
equation
$$
{\bf J}^n\cdot\delta{\bf y}^n=-{\bf F}^n. \eqno(A4)
$$
In (A4) ${\bf J}$ is the Jacobian matrix,
$J_{lm}\equiv\partial F_l/\partial y_m$.
The parameter $\omega$ should be chosen to optimize the convergency of the
iteration process to a solution. We use the following form of $\omega$,
$$
\omega={\eta\over max(\eta,\Delta)}, \eqno(A5)
$$
where the parameter $\eta=0.03$ and $\Delta$ is the average relative
correction,
$$
\Delta={1\over K}\sum_{k=1}^K\left|{\delta y_k\over y_k}\right|.
$$

In the case of equations (17), (18) we have
$K=2(I-1)$ functional relations involved $2I$
variables, $v_i$ and $(c^2_s)_i$. Two variables, $v_I$ and $(c_s)_I$,
must be fixed as
boundary conditions, when the number of independent variables
equals to the number of equations, and system (A2) can be solved.
In the case of equations (26)-(28) we have $K=3(I-1)$ and
three boundary values, $v_I$, $(c_s)_I$ and $\Omega_I$, which are fixed to
provide a consistent solution of (A2).

The presence of the singular points $(r_s)_{in}$ which coincides with
$r_1$ introduces
additional complications to
our method. To satisfy to two regularity conditions,
$$
D=0 \quad{\rm and}\quad N=0, \eqno(A6)
$$
at $r_1$ we modify the iteration procedure by the following way.
We add the equation $D=0$ to the set of equations (A1) together with a new
independent variable $\ell_{in}$. Applying the Newton-Raphson iteration
scheme we obtain a solution in which $D=0$ at $r_1$, but
$N\neq 0$ in general. To satisfy the condition $N=0$ at the point $i=1$
the appropriate
choice of $r_1$ must be done. We find the correct value of $r_1$
using the bisection method in which $r_1$ is changed by displacing
all the grid.
We apply the Newton-Raphson iterations (A4) for each new grid location.

To correctly approximate the differential equations in the first grid
interval ($r_1,r_2$), where $r_1=(r_s)_{in}$,
we expand solution at $r_1$,
$$
v(r)=v_1+v_s'(r-r_1), \eqno(A7)
$$
$$
c_s^2(r)=(c_s^2)_1+(c_s^2)_s'(r-r_1). \eqno(A8)
$$
The procedure of calculation of the coefficients
$v_s'$ and $(c_s^2)_s'$ are given in Appendix~B.
Using (A7) and (A8) we fix the values at the second grid point as follows,
$$
v_2=v_1+v_s'(r_2-r_1), \eqno(A9)
$$
$$
(c_s^2)_2=(c_s^2)_1+(c_s^2)_s'(r_2-r_1). \eqno(A10)
$$
Relations (A9) and (A10) are used instead of the correspondent difference
equations in (A1).
The modified system of difference equations avoids numerical instabilities
connected with the presence of the inner singular point and allows us
to obtain a solution crossed continuously this point.

We have found that described numerical procedure becomes unstable
in the case of equations (26)-(28) at large values of $\alpha$ and $\dot{m}$.
The numerical instability arises because of influence of equation (26)
and results in small-scale oscillations of all quantities.
The numerical instability of similar type was found by Beloborodov (1998),
who suppressed the oscillations on the level of $\sim 10^{-3}$ of relative
amplitude by applying a smoothing procedure to $\Omega$.
Unfortunately, such a smoothing procedure
can not be included into our method because
of additional coupling of equations (A1) with the regularity conditions
(A6).

\section{Expansion at singular point}

We consider first the case of equations (26)-(28).
We expand the numerator $N_2$ and denominator $D_2$ at the singular
point $r_s$, as follows
$$
N_2(r)=\left({\partial N_2\over\partial r}+{\partial N_2\over\partial v}v_s'+
{\partial N_2\over\partial c_s^2}(c_s^2)_s'+{\partial N_2\over\partial
\Omega}\Omega_s' \right)(r-r_s), \eqno(B1)
$$
$$
D_2(r)=\left({\partial D_2\over\partial r}+{\partial D_2\over\partial v}v'+
{\partial D_2\over\partial c_s^2}(c_s^2)' \right)(r-r_s). \eqno(B2)
$$
In (B1) and (B2) the partial derivatives are taken at $r=r_s$,
$v=v_s$, $c_s^2=(c_s^2)_s$ and $\Omega=\Omega_s$, and
we use notations $v_s=v(r_s)$, $(c_s^2)_s=c_s^2(r_s)$
and $\Omega_s=\Omega(r_s)$.
The `prime' mark means the radial derivative of the correspondent quantity
in (B1) and (B2).
In (B2) we take into account that $\partial D_2/\partial\Omega=0$.
We denote for convenience,
$$
\xi_s=r_s{v_s'\over v_s}, \qquad \eta_s=r_s{(c_s^2)_s'\over (c_s^2)_s}, \qquad
\chi_s=r_s{\Omega_s'\over\Omega_s}.
$$
The value of $\chi_s$ can be defined with help of (26).
Substituting (B1) and (B2) into equations (27) and (28) we finally obtain
a quadratic equation with respect to $\xi_s$:
$$
\left(v_s{\partial D_2\over\partial v}+2(c_s^2)_s
{\partial D_2\over\partial c_s^2}A\right)\xi_s^2+
\left[r_s{\partial D_2\over\partial r}+2(c_s^2)_s\left(
{\partial D_2\over\partial c_s^2}B-{\partial N_2\over\partial c_s^2}A\right)-
v_s{\partial N_2\over\partial v}\right]\xi_s-
$$
$$
\qquad\qquad
\left(r_s{\partial N_2\over\partial r}+\Omega_s{\partial N_2\over\partial
\Omega}\chi_s+2(c_s^2)_s{\partial N_2\over\partial c_s^2}B\right)=0,
\eqno(B3)
$$
where we denote
$$
A=1-{\cal M}_s^2, \quad {\rm and} \quad B={3\over 2}+{r_s\over r_s-2}+
{1\over (c_s^2)_s}\left(
\Omega_s^2 r_s^2-{r_s\over (r_s-2)^2}\right).
$$
Having $\xi_s$ from (B3) one obtain $\eta_s=A\xi_s+B$.
The found values of $\xi_s$, $\eta_s$ and $\chi_s$ determine derivatives
$v_s'$, $(c_s^2)_s'$ and $\Omega_s'$ which we use
in the numerical procedure discussed in Appendix~A.
Equation (B3) has two roots. Which root should be used is defined by
convergency of the iterations.
We have found that only one of the two roots corresponds to the
convergent solution.
The partial derivatives of $N_2$ and $D_2$ used in (B3) can be calculated
analytically or using numerical differentiation from (29) and (30).
The latter method is simpler, but less accurate than the analytic one.
The analytic derivation requires some efforts and produces quite long
expressions which we do not present here.
In the numerical procedure we have used the analytical expressions for
derivatives, but in addition, we have checked them using estimates from
the numerical differentiation.

To judge the type of critical point we follow the procedure
described by Kato et al. (1998). We introduce a new variable $\tau$
defined by
$$
d\tau={dr\over rD_2}. \eqno(B4)
$$
From equation (26)-(28) one obtain
$$
{d\Omega\over d\tau}=\chi\Omega D_2, \quad
{dv\over d\tau}=v N_2, \quad
{dc_s^2\over d\tau}=2c_s^2(AN_2+BD_2). \eqno(B5)
$$
All of the variables are now expanded around those at $r_s$,
$$
r=r_s+\Delta r, \quad v=v_s+\Delta v, \quad c_s^2=(c_s^2)_s+\Delta c_s^2,
\quad \Omega=\Omega_s+\Delta\Omega. \eqno(B6)
$$
Substituting (B6) into (B4) and (B5) and retaining only linear terms,
one obtain a system of linear differential equations
with respect to $\Delta r$, $\Delta v$, $\Delta c_s^2$ and $\Delta\Omega$.
Assuming that these quantities depend on $\tau$ in the form
$\exp(\lambda\tau)$, one obtain the following characteristic equation
which determines the eigenvalues $\lambda$:
$$
\lambda^2-\lambda\left[r_s{\partial D_2\over\partial r}+
v_s{\partial N_2\over\partial v}+2(c_s^2)_s\left(
A{\partial N_2\over\partial c_s^2}+
B{\partial D_2\over\partial c_s^2}\right)\right]-
$$
$$
\qquad
2 A r_s(c_s^2)_s\left({\partial N_2\over\partial r}
{\partial D_2\over\partial c_s^2}-{\partial D_2\over\partial r}
{\partial N_2\over\partial c_s^2}+{\partial D_2\over\partial c_s^2}
{\chi_s\over r_s}\Omega_s{\partial N_2\over\partial\Omega}\right)-
$$
$$
\qquad
2 B v_s(c_s^2)_s\left({\partial D_2\over\partial v}
{\partial N_2\over\partial c_s^2}-{\partial N_2\over\partial v}
{\partial D_2\over\partial c_s^2}\right)+
$$
$$
\qquad
r_sv_s\left({\partial D_2\over\partial r}{\partial N_2\over\partial v}-
{\partial N_2\over\partial r}{\partial D_2\over\partial v}\right)-
v_s{\partial D_2\over\partial v}
\chi_s\Omega_s{\partial N_2\over\partial\Omega}=0.
\eqno(B7)
$$
Again, all the partial derivatives in (B7) are taken at $r=r_s$,
$v=v_s$, $c_s^2=(c_s^2)_s$ and $\Omega=\Omega_s$.
If two solutions of quadratic equation (B7) are real and have different signs,
$\lambda_1\lambda_2<0$,
then the singular point is a {\it saddle}. The point is a {\it node} if
two real roots of (B7) have identical signs,
$\lambda_1\lambda_2>0$. Complex conjugate roots of (B7) corresponds to the
{\it spiral} singular point.

In the case of equations (17), (18) equations (B3) and (B7)
must be modified by
substituting $N_1$ and $D_1$ instead of $N_2$ and $D_2$, and assuming
$\partial N_1/\partial\Omega=0$.
The notation for $B$ must be also changed to
$$
B={3\over 2}+{r_s\over r_s-2}+
{1\over (c_s^2)_s}\left[\left({\ell_{in}\over r_s}+\alpha{(c_s^2)_s\over v_s}
\right)^2-
{r_s\over (r_s-2)^2}\right].
$$

\bigskip\bigskip
{
\footnotesize
\StartRef
\noindent {\large \bf References} \\

\Ref Abramowicz, M. A., \& Zurek, W. H. 1981, ApJ, 246, 314 \\

\Ref Abramowicz, M. A., \& Kato, S. 1989, ApJ, 336, 304 \\

\Ref Abramowicz, M. A., Czerny, B., Lasota, J. P., \& Szuszkiewicz, E.
1988, ApJ, 332, 646 \\

\Ref Beloborodov, A. M. 1998, MNRAS, 297, 739 \\

\Ref Chakrabarti, S.K. 1996, ApJ, 464, 664 \\

\Ref Chakrabarti, S.K., \& Molteni, D. 1993, ApJ, 417, 671 \\

\Ref Chen, X., \& Taam, R. E. 1993, ApJ, 412, 254 \\

\Ref Frank, J., King, A., \& Raine, D. 1992, Accretion Power in Astrophysics
(Cambridge Univ. Press) \\

\Ref Fukue, J. 1987, PASJ, 39, 309 \\

\Ref H\"oshi, R., \& Shibazaki, N. 1977, Prog. Theor. Phys., 58, 1759 \\

\Ref Igumenshchev, I. V., Abramowicz, M. A., \& Novikov, I. D. 1998,
MNRAS, 298, 1069 \\

\Ref Kato, S., Fukue, J., \& Mineshige, S. 1998, Black-Hole Accretion Disks
(Kyoto: Kyoto Univ. Press) \\

\Ref Liang, E. P. T., \& Thompson, K. A. 1980, ApJ, 240, 271 \\

\Ref Matsumoto, R., Kato, S., Fukue, J., \& Okazaki, A. T. 1984,
PASJ, 36, 71 \\

\Ref Muchotrzeb, B., \& Paczy\'nski, B. 1982, Acta Astr., 32, 1 \\

\Ref Muchotrzeb, B. 1983, Acta Astr., 33, 79 \\

\Ref Muchotrzeb-Czerny, B. 1986, Acta Astr., 36, 1 \\

\Ref Narayan, R., Kato, S., \& Honma, F. 1997, ApJ, 476, 49 \\

\Ref Novikov, I. D., \& Thorne, K. S. 1973, in
C. DeWitt \& B. DeWitt, eds, Black Holes.
Gordon \& Breach, New York, 343 \\

\Ref Paczy\'nski, B., \& Wiita, P. J. 1980, A\&A, 88, 23 \\

\Ref Paczy\'nski, B., \& Bisnovatyi-Kogan, G.S. 1981, Acta Astr., 31, 283 \\

\Ref Pringle, J. E., Rees, M. J., \& Pacholczyk, A. G. 1973, A\&A, 29, 179 \\

\Ref Pringle, J. E. 1981, ARA\&A, 19, 137 \\

\Ref Shakura, N.I. 1972, Astron. Zh., 49, 921 \\

\Ref Shakura, N.I., \& Sunyaev, R.A. 1973, A\&A, 24, 337 \\

\newpage
\vskip 5in
\newpage

\begin{figure}
\plottwo{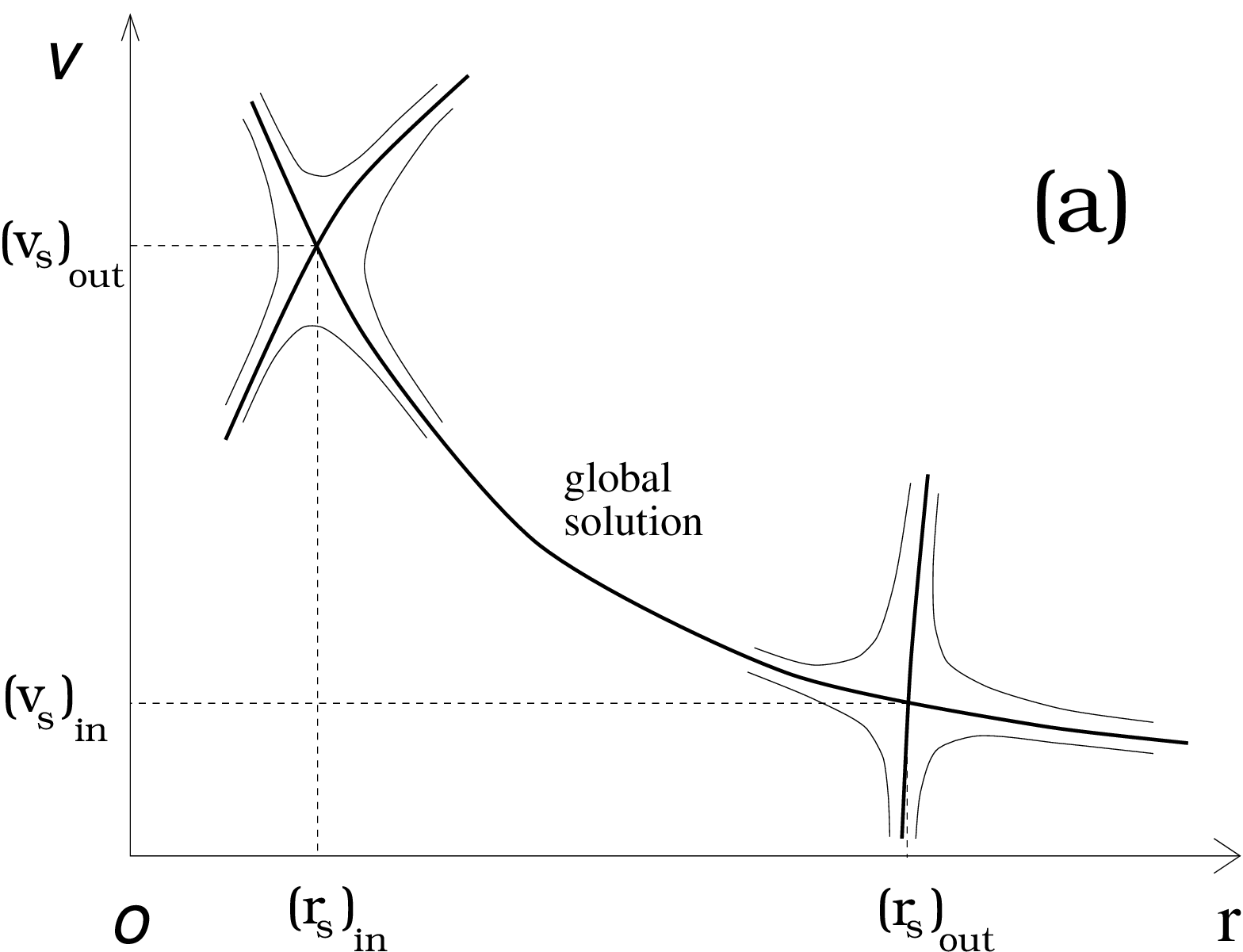}{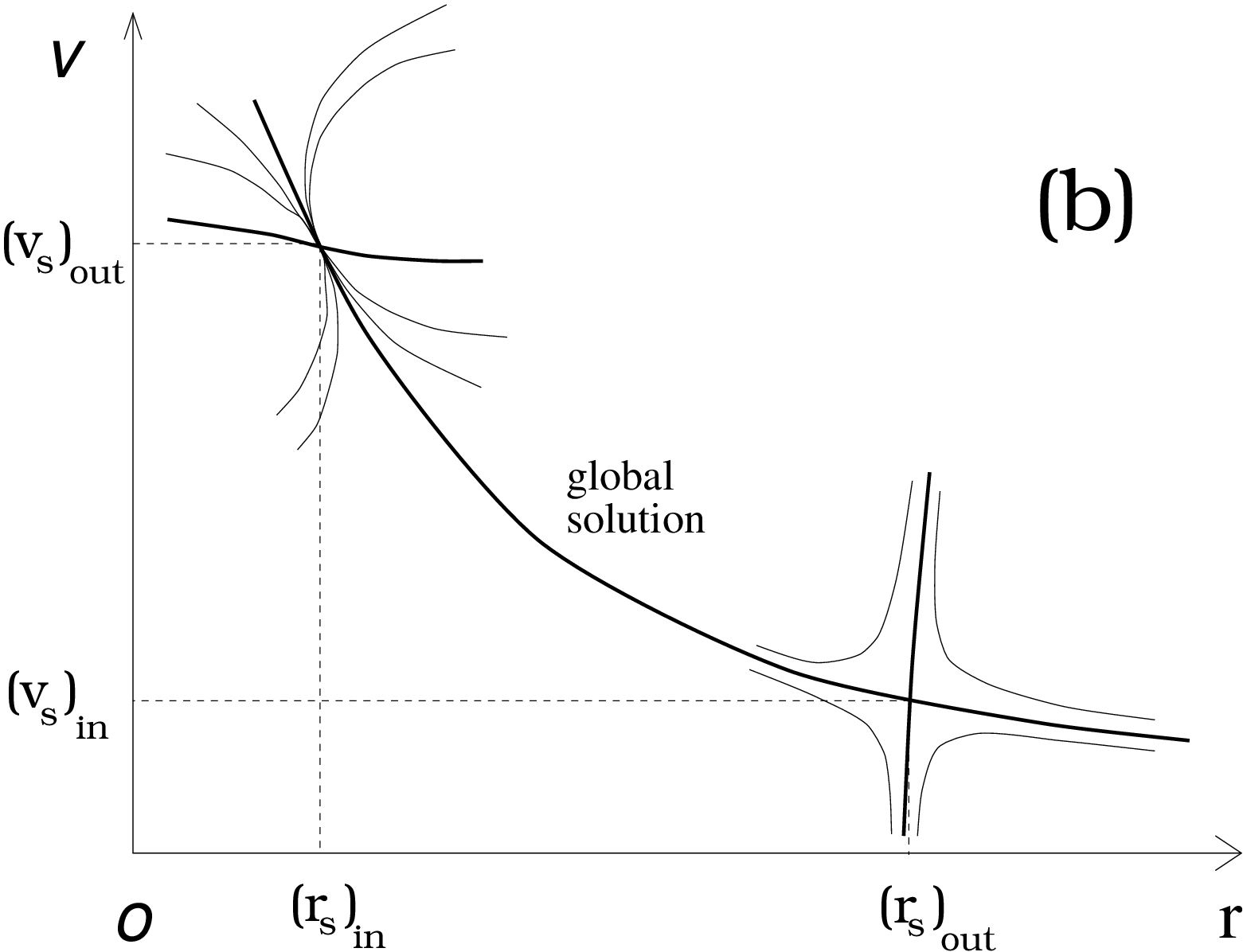}
\caption{
Schematic illustration of behaviour of the integral curves
in the vicinity of a global solution for transonic accretion disks.
Two singular points exist at $r=(r_s)_{in}$ and $(r_s)_{out}$ in the case of
viscosity prescription (1).
The outer singular point at $(r_s)_{out}$ is always of a saddle type.
The inner singular point at $(r_s)_{in}$ locates close to the last stable
black hole orbit at $6 r_g$ and
changes its type from a saddle to nodal one with increase of
$\alpha$ as shown on panels (a) and (b), respectively.
Only the separatrix (thick line) which passes through both singular points
represents the global solution. The nodal type inner singular point
does not mean the existence of multiple physical solutions.
\label{fig1}}
\end{figure}

\begin{figure}
\plottwo{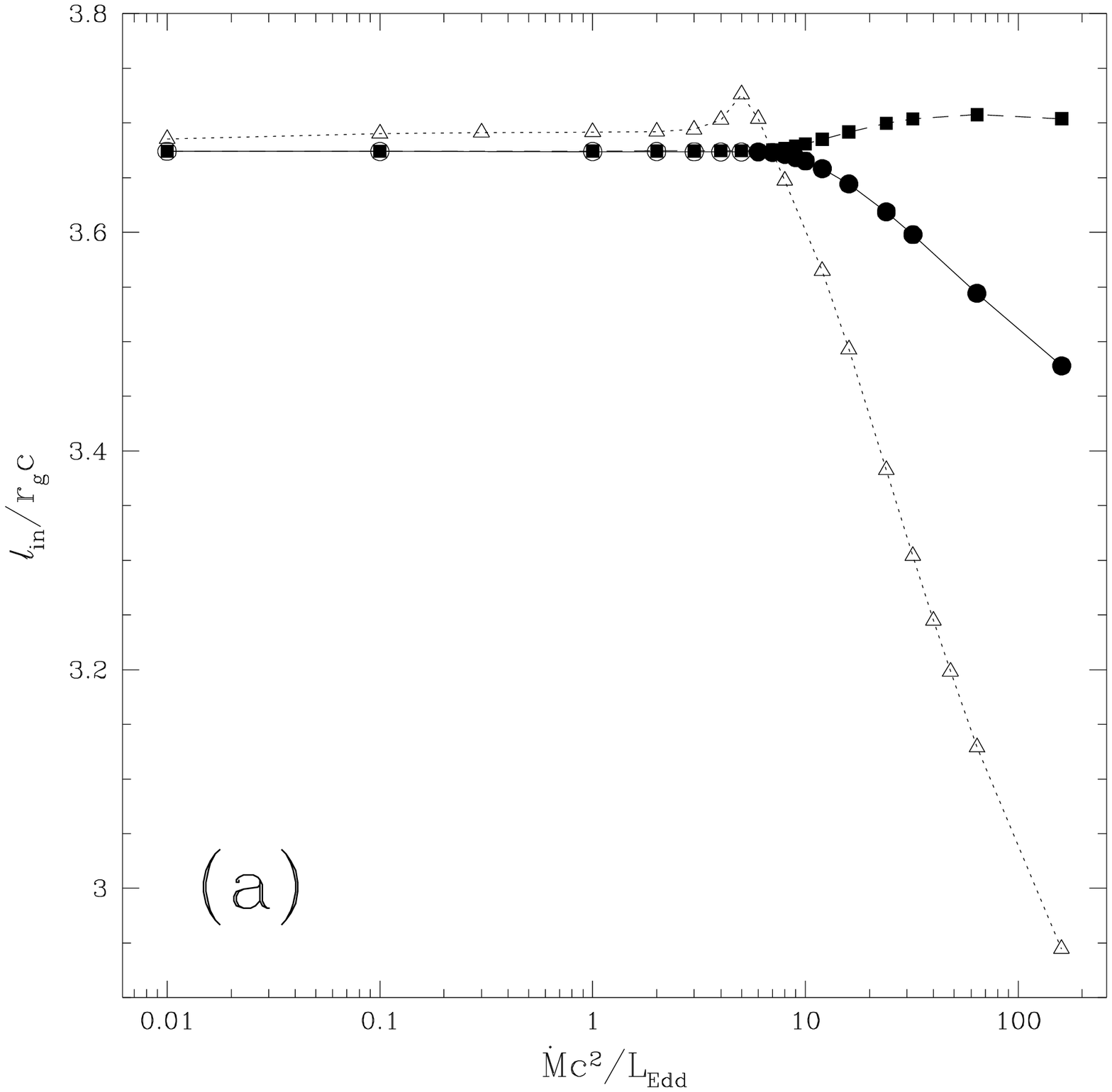}{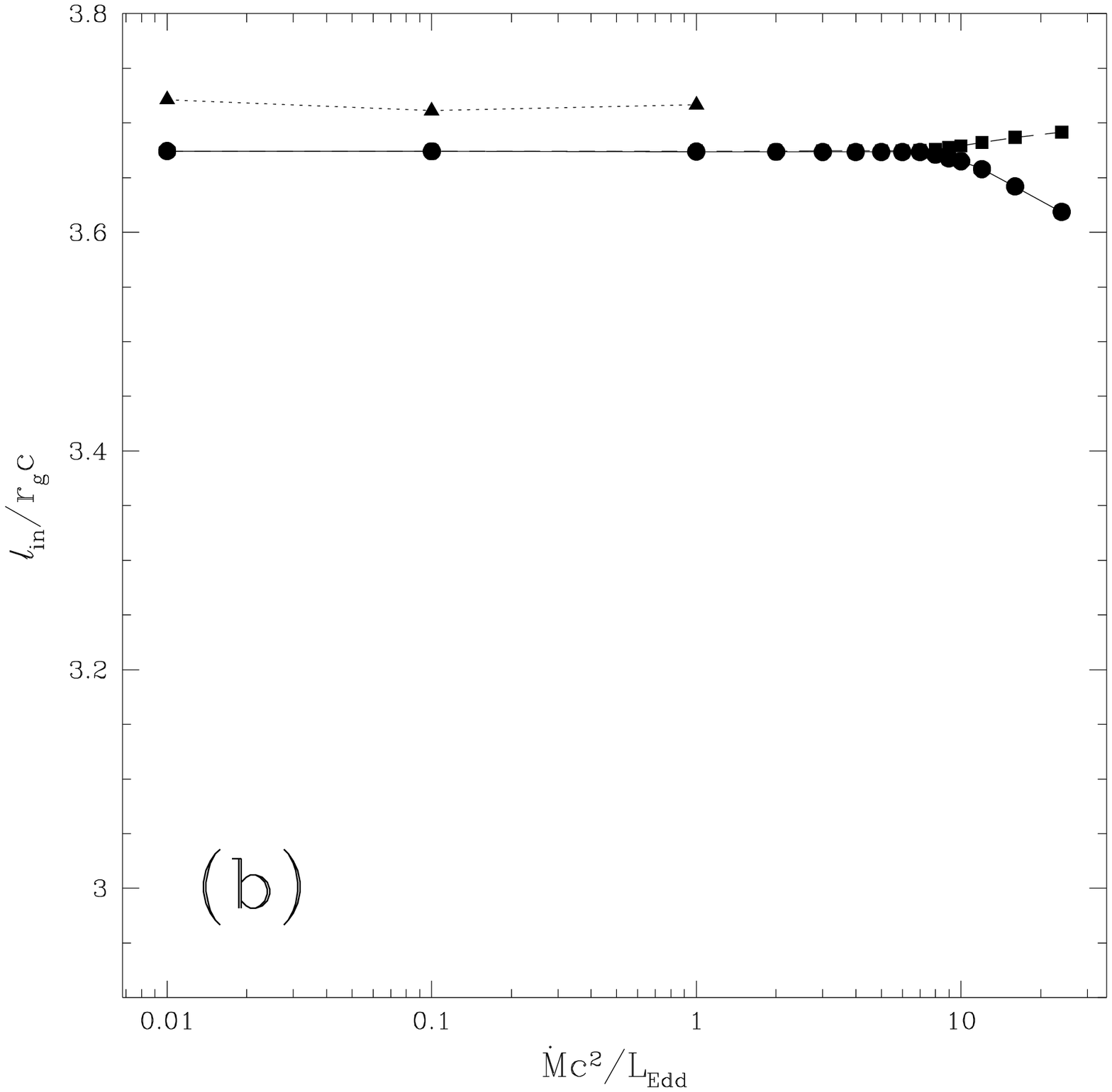}

\plottwo{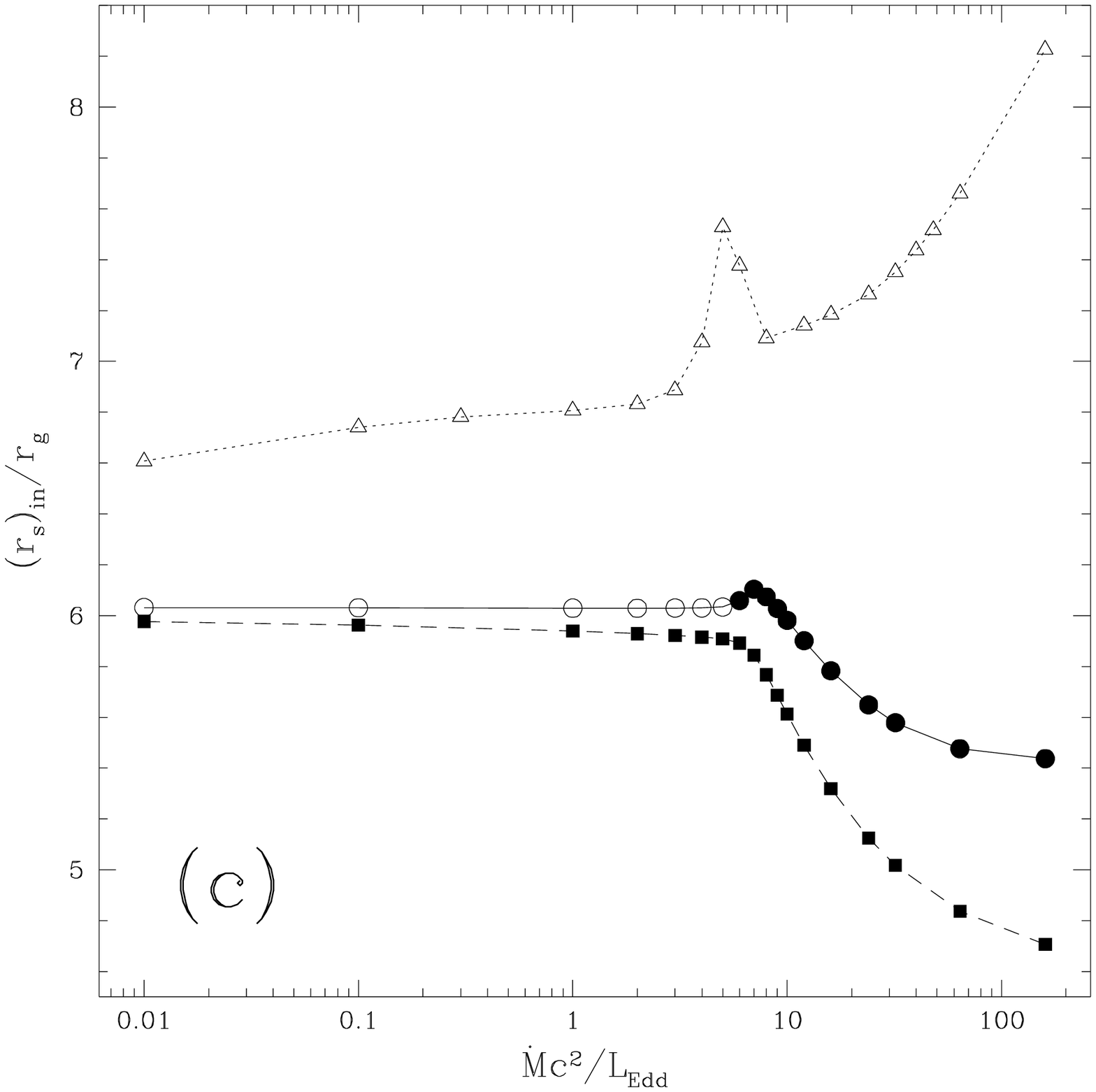}{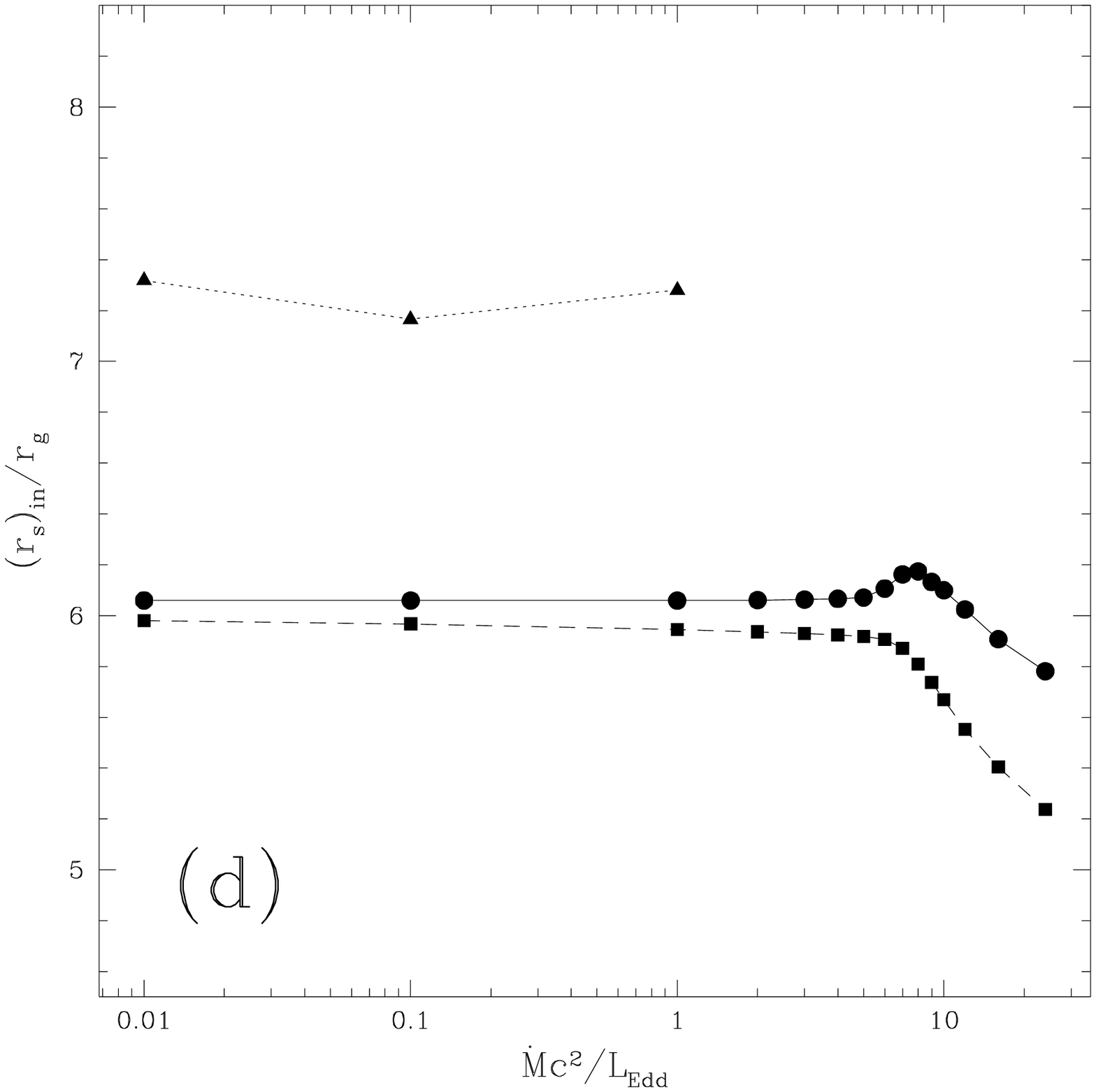}
\caption{The specific angular momentum $\ell_{in}$
[panels (a) and (b)] and the position of the inner singular points
[panels (c) and (d)]
as a function of the mass accretion rate $\dot{M}$
for different viscosity parameters
$\alpha=0.01$ (squares), $0.1$ (circles) and $0.5$ (triangles).
Panels (a) and (c) correspond to viscosity prescriptions (1)
and panels (b) and (d) correspond to viscosity prescriptions (2).
The solid dots represent models with the saddle-type
inner singular points, whereas the empty dots correspond to the nodal-type ones.
\label{fig2}}
\end{figure}

\begin{figure}
\plotone{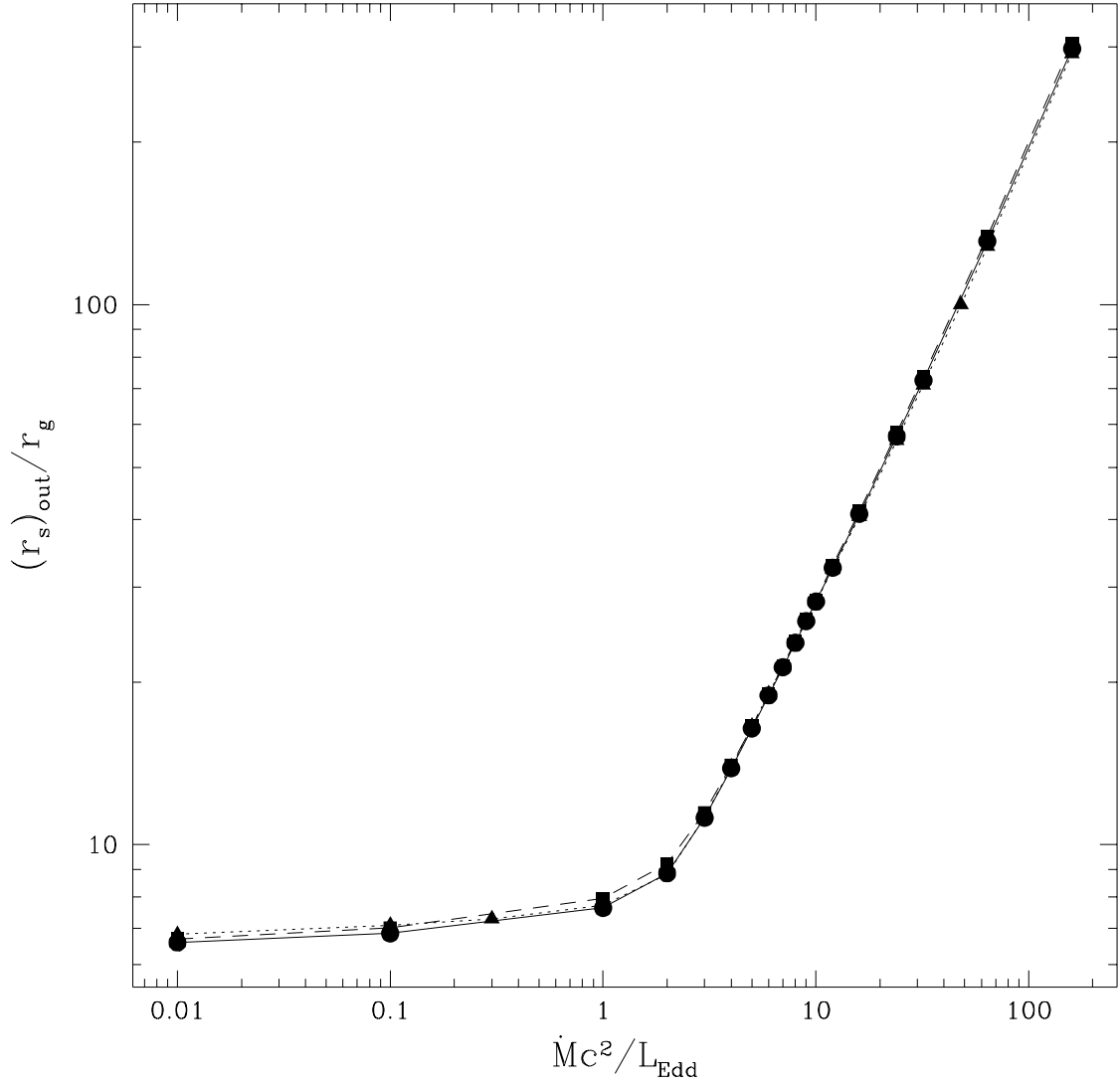}
\caption{Location of the outer singular points as a function of
the mass accretion rate $\dot{M}$ in the case of viscosity prescription (1).
Models with $\alpha=0.01$, $0.1$ and $0.5$ are shown.
The points are always of a saddle-type.
See caption to Fig.2 for details of notations.
\label{fig3}}
\end{figure}

\begin{figure}
\plotone{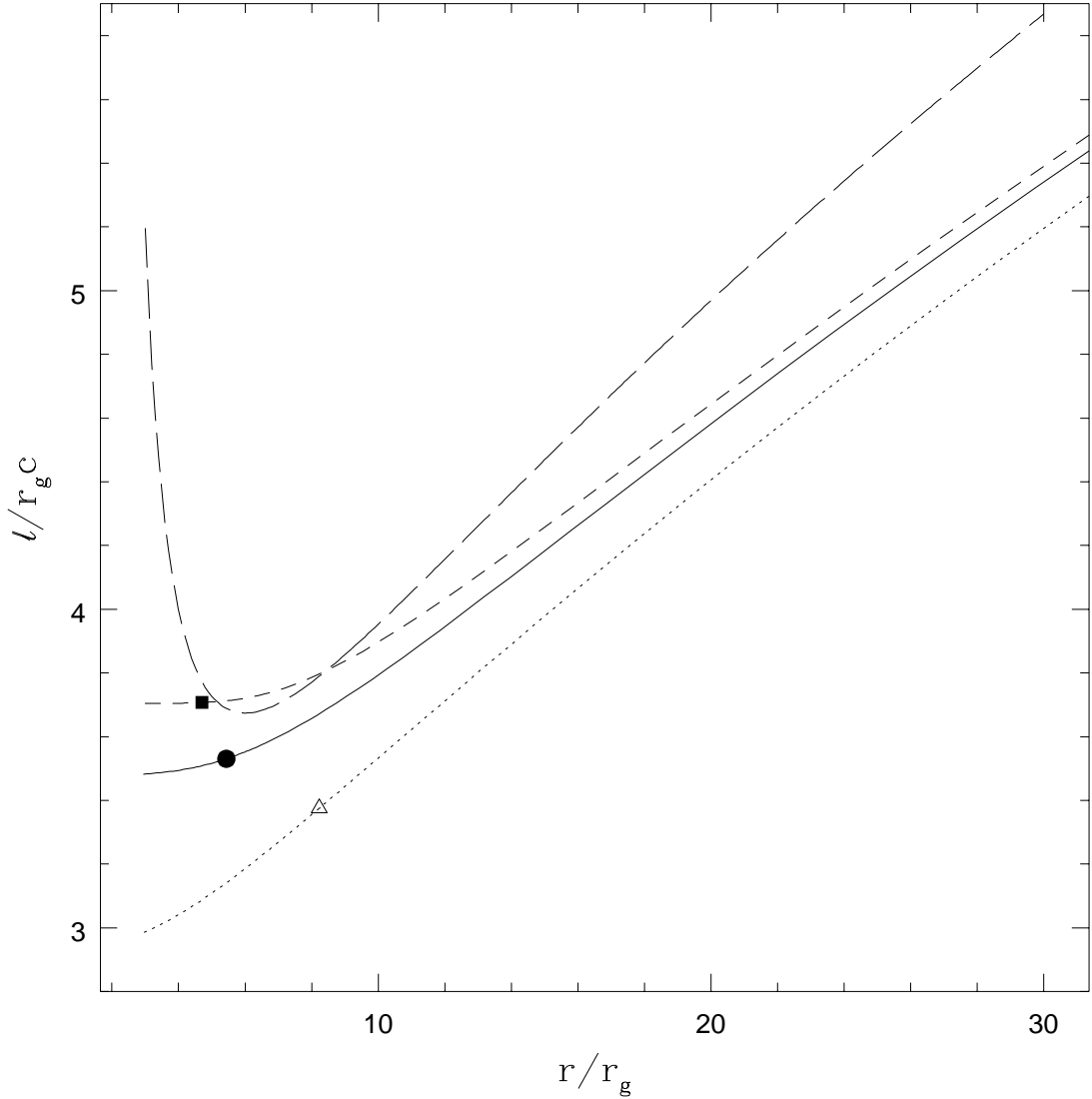}
\caption{The specific angular momentum distribution with respect to radius
in the innermost region of the disk with $\dot{m}=160$ and
viscosity prescription (1).
The long-dashed line corresponds to the Keplerian angular momentum.
The short-dashed, solid and dotted lines correspond to
$\alpha=0.01$, $0.1$ and $0.5$, respectively.
The dots on the curves show the position of the inner singular points
and the correspondent angular momentum. In the case of $\alpha=0.01$ and $0.1$
the singular points are of a saddle-type, and in the case of $\alpha=0.5$
the point is of a nodal-type.
\label{fig4}}
\end{figure}

\begin{figure}
\plottwo{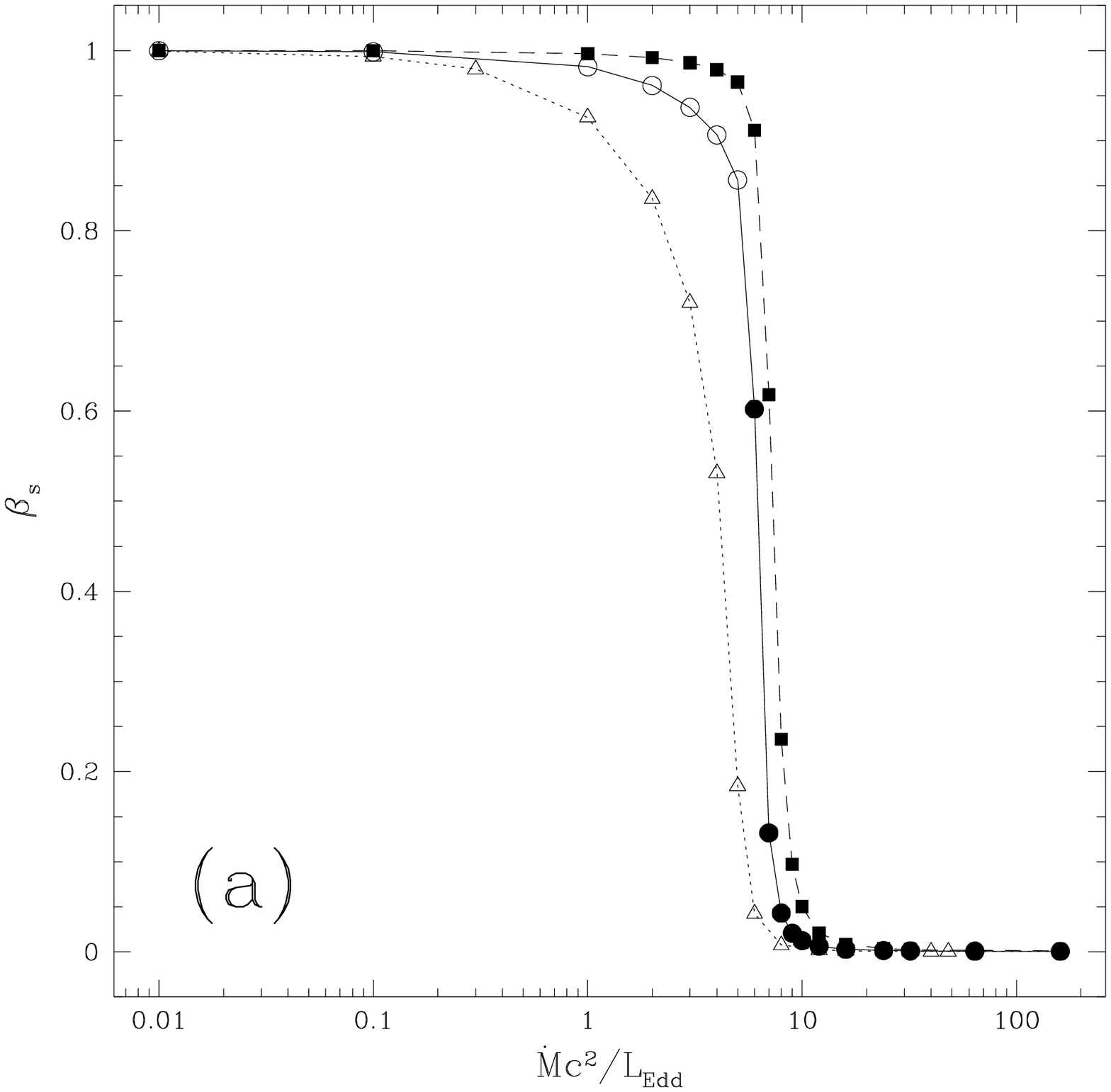}{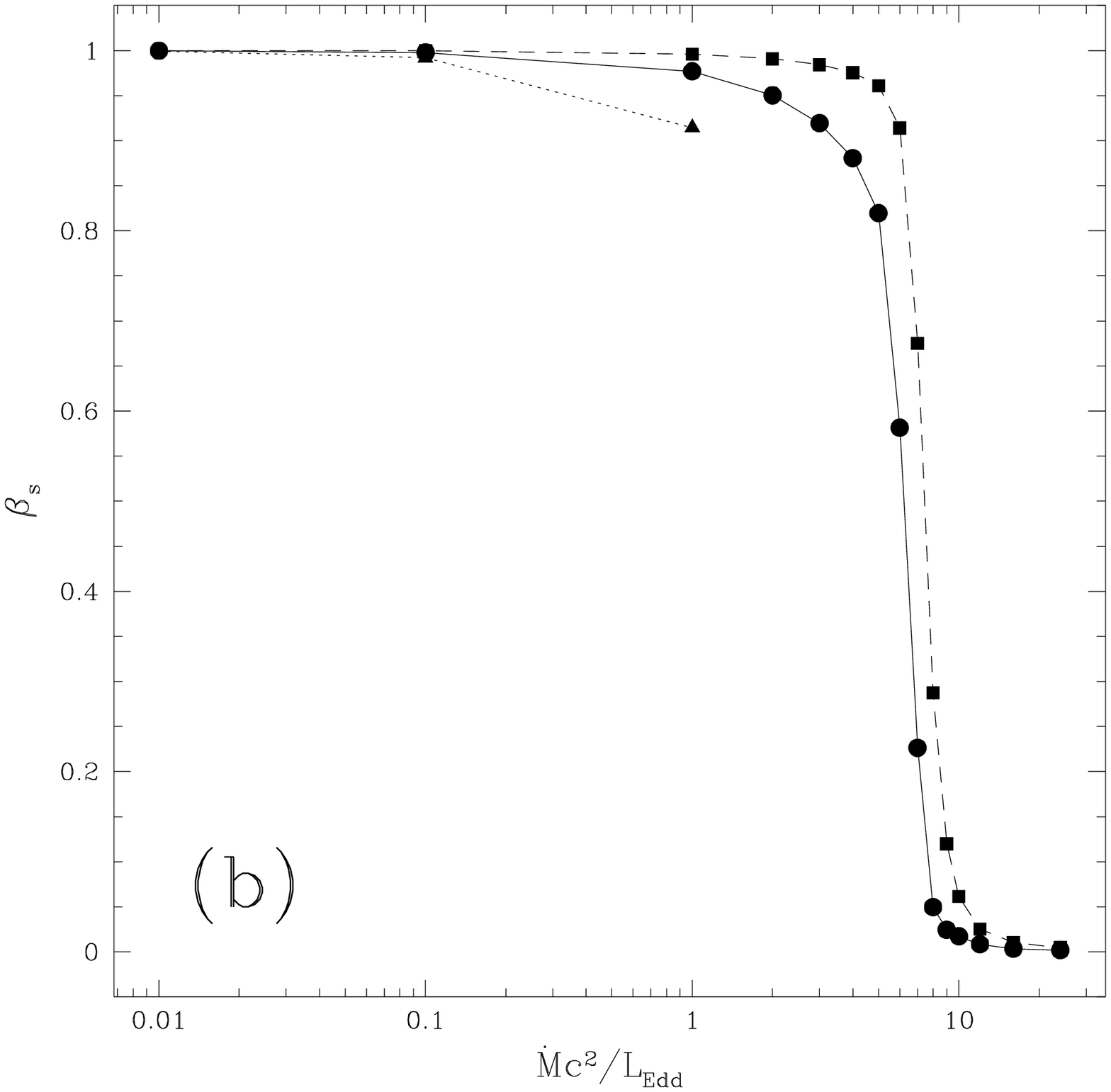}
\caption{Ratio of the gas pressure to the total pressure at
the inner singular point, $\beta_s$,
as a function of the mass accretion rate $\dot{M}$.
Panels (a) and (b) correspond to viscosity prescriptions (1) and (2),
respectively. Models with $\alpha=0.01$, $0.1$ and $0.5$ are shown.
See caption to Fig.2 for details of notations.
\label{fig5}}
\end{figure}

\begin{figure}
\plotone{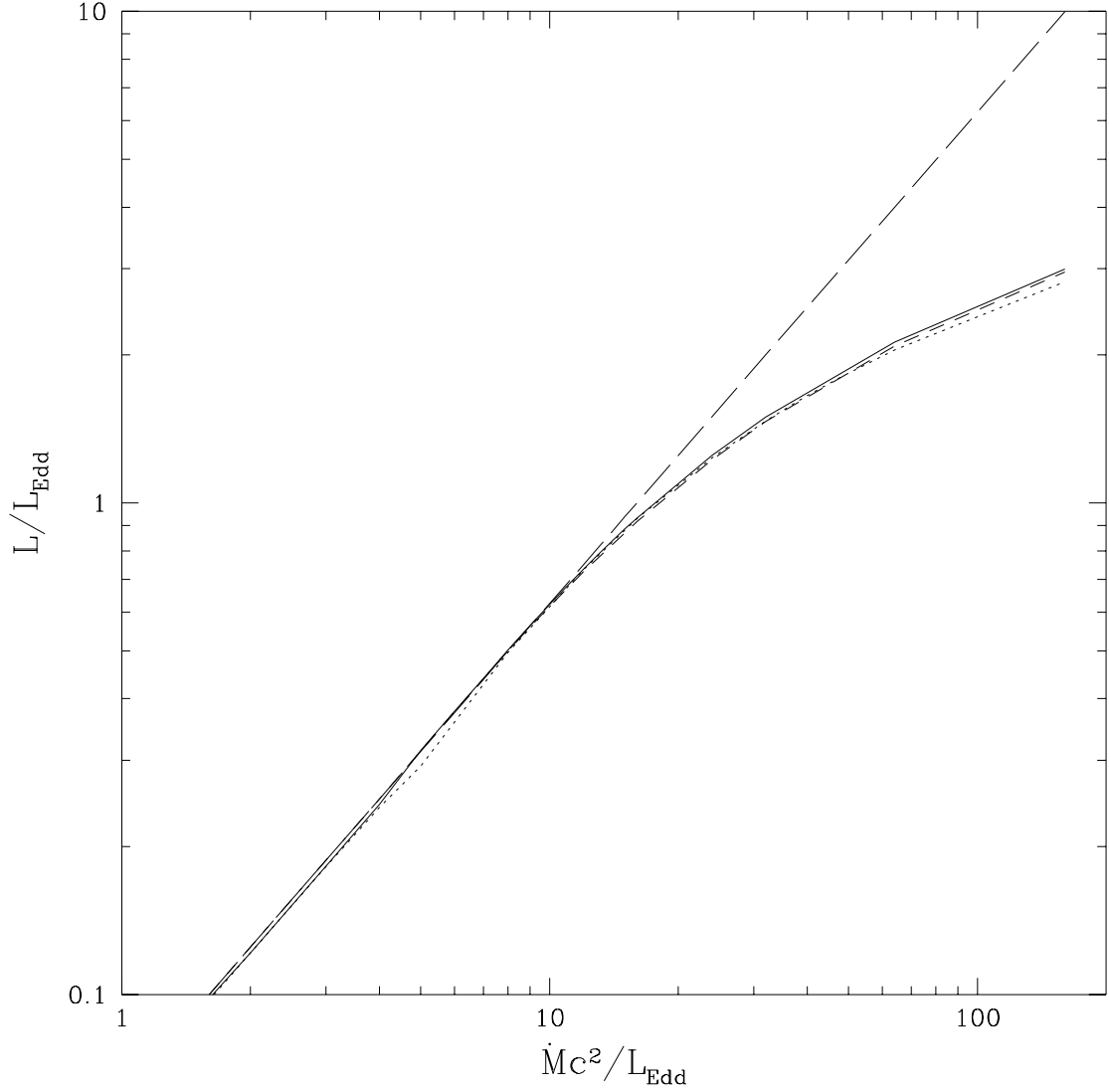}
\caption{Luminosities of accretion disks at different values of $\dot{M}$ and
$\alpha$ in the case of viscosity prescription (1). 
The long-dashed line corresponds to the standard model with
the radiative efficiency $1/16$. Deviation of luminosities
from one for the standard model indicates importance of advection.
The short-dashed, solid and dotted lines correspond to $\alpha=0.01$,
$0.1$ and $0.5$, respectively.
\label{fig6}}
\end{figure}

\end{document}